# Thermodynamics and Kinetics of Silicate Glasses submitted to Binary Ion Exchange: Equilibrium Conditions and Interdiffusion Kinetics

Guglielmo Macrelli[1]

[1]*Isoclima SpA – R&D Department*

*Via A.Volta 14, 35042 Este (PD), Italy*

*guglielmomacrelli@hotmail.com*

*Correspondence: guglielmomacrelli@hotmail.com

**Abstract**

Ion exchange processes between an ion reservoir and a solid matrix are modeled under the assumption that near interface volumes reach equilibrium in a much faster time than the overall ion exchange process time while, in the bulk of the solid matrix, ions are transported by an interdiffusion kinetic process. Ion exchange equilibrium conditions are initially established according to classical thermodynamics. The result is defined in terms of the chemical potentials of exchanging ionic species. The proposed original derivation is performed making use of the thermodynamics of subsystems to determine near-surface equilibrium concentrations and ion exchange isotherms. Interaction energies of the exchanging ions in the glass are determined through the thermodynamic factor *n*, which is a parameter of the ion exchange isotherm. The kinetics of the ion exchange process in silicate glass is discussed to find connections with the near-surface equilibrium condition. The flux equation for the incoming ions in the glass is written in the form of a Fick equation with a concentration-dependent interdiffusion coefficient incorporating the thermodynamic factor *n*. It has been found that the thermodynamic factor of the interdiffusion coefficient is related to the interaction energy of the exchanging ions in the glass allowing a new approach to the interpretation of past experimental results. Surface concentration has been found substantially connected to the second parameter of the ion-exchange isotherm which is the equilibrium constant *K* and significativetly influenced by the chemical composition of the ion reservoir (the molten salt bath). Further treatment of kinetics of ion-exchange within the framework of non-equilibrium thermodynamics allows the interpretation of the thermodynamic factor as a function of incoming ions concentration in the glass matrix.

## 1. Introduction

The first sentence of the introduction of the classic book of Helfferich[1] about Ion Exchange is an old statement in Latin language: "corpora non agunt nisi fluida sive soluta" that is: "substances don't react unless in liquid or dissolved state". This old statement can be dated back to Aristotle[2] and it can be understood as an empirical rule which has driven most part of the alchemic efforts to the search of a universal solvent[3]. Even though it has been mitigated by Hedvall[4] considering its original Greek language formulation, this approach has eventually driven ancient alchemists to some questionable positions[3]. As correctly affirmed by Helfferich[1], Ion Exchange is a chemical reaction not respecting the above "ancient rule". Most part of modern solid state chemistry[5] and electrochemistry[6] clearly indicate that the above mentioned ancient rule is not applicable. Ion Exchange is a chemical reaction between two materials or two phases or volumes of the same material where ions in the two media are exchanged by the effect of a driving force. It is not limited to the state or the phases of materials. In essence it is a kinetic process driven to achieve thermodynamic equilibrium conditions. When, during ion exchange, dissipative effects occur like stress build-up and relaxation or structural changes in the ion exchange hosting materials it can be categorized in the class of non-equilibrium irreversible processes. The only limiting conditions to ion exchange are the laws of physics: thermodynamics and kinetics. The history of Ion Exchange can be dated back to mentions in the ancient literature[7,8] through evolutions embracing: soil geochemistry, waters treatments, chromatography, application to adsorption of fission products. In this study theoretical conditions and applications of ion exchange to silicate glasses are discussed. The scientific discovery of ion exchange effects in silicate glasses can be dated back to the last decades of the past century[9,10,11]. Since that point in time, Ion Exchange evolved from laboratory conditions to applications in glass industrial products with significant technical and commercial impact. The scientific understanding of Ion Exchange in silicate glass is established in the literature[12,13,14,15] even though some residual issues about Ion Exchange and related effects in silicate glass are still open[16,17], this has not prevented the spread of technology in a wide number of applications. In essence the process is understood as an interdiffusion kinetic process driven by the gradients of the electrochemical potentials of the involved ions. It is a binary chemical reaction between the glass matrix and a reservoir of ions usually consisting in a molten salts bath. Based on this last definition it is clearly a kinetic process that will evolve up to the point where equilibrium conditions are achieved. Classical thermodynamics is a science of equilibrium, hence the analysis of Ion Exchange from the thermodynamic point of view can only define the conditions to which the process will eventually evolve while, to understand the evolution in time, a kinetic analysis is requested. An interesting approach is to separate the process in two portions of the glass articles: near surface and bulk. This separation is useful in assuming that the time to equilibrium at the interface between glass and reservoir is much smaller than the overall time of the process. This last assumption allows to define a boundary condition to the kinetic interdiffusion process in establishing a constant equilibrium surface concentration achieved almost instantaneously as the glass and the reservoir get in contact. This last assumption is the basis to achieve the most popular solution to the interdiffusion kinetic equation for the concentration of the incoming ions consisting in the so-called "complementary error function" (*erfc*) solution[14,18]. The main purpose of this work is to establish the scientific fundamentals of both equilibrium and non equilibrium science of ion exchange in silicate glasses. The two aspects are connected and one of the purposes is to clarify connections and assumptions. A theoretical analysis is provided for the equilibrium conditions and, on that basis, a critical review of past literature is performed where experimental methods and results are analyzed. In the second part of this study the non-equilibrium kinetic description of binary ion exchange is proposed based on the non-equilibrium thermodynamics of irreversible processes.

Ion Exchange is a widely used chemical process to increase mechanical strength of silicate glasses[9,14] or to create optical waveguides by modifying the near surface refractive index[10,20]. The process is typically performed by putting the glass in contact with a ions reservoir made of a molten salt bath. The entire process can be modeled in two parts as depicted in Figure 1: a surface equilibrium process where the time to reach an equilibrium condition is far lower that the entire process time and an interdiffusion kinetic process in the bulk of the solid glass matrix[19].

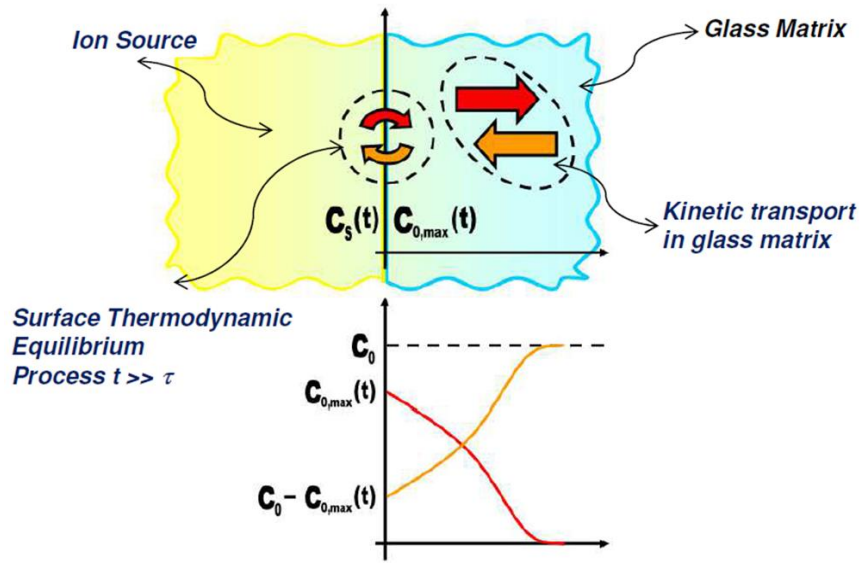

Figure 1 – Model of the Ion Exchange Process in the solid glass.

The equilibrium conditions are important to define the boundary condition for the interdiffusion kinetic process[19]. In this study equilibrium condition are derived making use of classical thermodynamics. Even though its fundamentals (the first, the second and the zeroth laws) have been established in the past couple of centuries it is probably still one of the most discussed subjects of physics. Looking to the books recently published on "Classical Thermodynamics", either at undergraduate and graduate levels, one shall conclude that this theory is still offering interesting contribution to science when analysed from different viewpoints. It is worth mentioning a famous statement of Einstein about classical thermodynamics[21]: "*It is the only physical theory of universal content. I am convinced that, within the framework of applicability of its basic concepts, it will be never overthrown*". In this study equilibrium conditions for binary ion exchange are discussed. This discussion is based within the conditions for equilibrium, either thermal, mechanical and chemical, that can be classically found in the treatments of Kittel[22], Callen[23] and Kondepudy and Prigogine[24]. Those treatments consider subsystems[22,23] or discrete systems[24] as part of a total insulated system.

Binary ion exchange is a specific mass transfer interdiffusion process presenting additional constraints to both extensive and intensive parameters in comparison to simple diffusive mass transfer processes. The classical treatment of equilibrium conditions for mass transfer diffusion[22,23,24] does not consider binary ion exchange. The conclusion that diffusion equilibrium is achieved when the chemical potential of a single diffusion component is the same in the two subsystems cannot be generalized to ion exchange by assuming that chemical potentials of the two exchanging components shall be individually equal for the two subsystems. In this study it will be provided the correct equilibrium conditions for ion exchange. This result has already been established by Araujo[25] making use of potential thermodynamic functions constrained in extensive and intensive parameters. Herewith a new derivation is offered based on the classical approach of thermodynamic of sub systems[26]. The process of achieving equilibrium condition is further discussed in terms of kinetic making use of the concepts of non-equilibrium thermodynamics[18,27]. A connection between thermodynamics and kinetics is established recognizing that the "$n$" factor of the equilibrium isotherm is the thermodynamic factor of the interdiffusion coefficient. Past experimental results of the interdiffusion coefficient for a Potassium/Sodium ion exchange in soda-lime glass are analyzed providing a new interpretation based on the connection between surface thermodynamics equilibrium and interdiffusion kinetics.

### 2. Thermodynamics of sub systems: equilibrium conditions

Ion Exchange is defined as a binary chemical reaction between two subsystems in mutual contact that we will identify as 1 and 2. The two subsystems are considered open between each other and isothermal and isochoric. The sub systems 1 and 2 are parts of a total system that is considered insulated. The Ion Exchange reaction is binary because it involves a couple of ions: A and B which are resident in subsystem 1 and in subsystem 2. In order to discuss the specific conditions under which ions can be transferred between the subsystems, let's assume this may happen to achieve a thermodynamic equilibrium condition. In this case we can write the following equilibrium equation:

$$A + \bar{B} \rightleftarrows B + \bar{A}, \tag{1}$$

where A and B are two ions in Subsystem 1 and $\bar{A}$ and $\bar{B}$ are the same ions in Subsystem 2. Macroscopic equilibrium conditions for a system are usually set by the laws of thermodynamics. The ion exchange reaction indicated in equation (1) is constrained by two conditions: mass conservation and electroneutrality. An exemplary way to establish equilibrium conditions for Ion Exchange is discussed by Araujo[25] based on the combined first and second laws of thermodynamics for an open system that can be written[26,28]:

$$dU \leq TdS - P\delta V + \sum_i \mu_i \delta N_i. \tag{2}$$

Equation (2) combines extensive properties of the system namely: $U$, internal energy of the system, $S$ entropy and $N_i$ number of moles of the i-th component of the system that can be transported from the system to the surroundings and from the surroundings to the system and $V$ volume of the system, with intensive properties: $T$ the temperature of the system, $P$ pressure and $\mu_i$ chemical potentials of the i-th components. The differential of functions indicated by "$d$" is to be considered an exact differential while differential indicated by "$\delta$" is not exact[28]. In equation (2) it is assumed that all work (mechanical work) is represented by the $PV$ expression. The already mentioned discussion of equilibrium conditions proposed by Araujo[25] is based on the consideration of a constrained isothermal, constant volume thermodynamic system with constraints applied on extensive variables $N_i$ and on their conjugate intensive variables $\mu_i$. In order to represent ion exchange reaction as indicated by equation (1) we assume that Subsystem 1 can exchange with Subsystem 2 two components A and B represented by the relative conjugated extensive and intensive variables: ($N_A, \mu_A$) and ($N_B, \mu_B$). The schematic of binary Ion Exchange between he two subsystems 1 and 2 is represented in Figure 2.

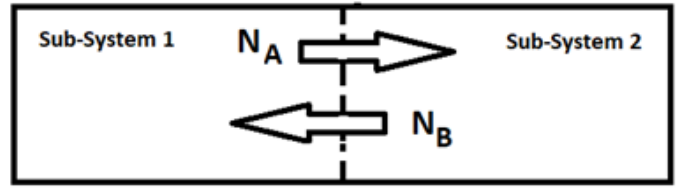

Figure 2 – Ion Exchange in the framework of thermodynamic of Sub-Systems.

Equilibrium condition is achieved[26] when the total entropy (that is the entropy of Subsystems 1 and 2, $S_T=S_1+S_2$) achieves its maximum:

$$\frac{\partial S_T}{\partial N_A} = 0. \tag{3}$$

This equilibrium condition is constrained by keeping constant the extensive variable $N_{total}=N_A+N_B$. The assumption for binary Ion Exchange is that exchanged components are conserved both in terms of exchanged components within the same subsystem and between the two subsystems (electroneutrality requirements), such conditions are set in the following equations:

$$(\delta N_A)_1 = -(\delta N_B)_1, \tag{i}$$

$$(\delta N_A)_2 = -(\delta N_B)_2, \tag{ii}$$

$$(\delta N_A)_1 = -(\delta N_A)_2. \tag{iii}$$

Now we take the assumption that the total system made of the two subsystems is an insulated system. In insulated systems internal energy is constant[26,28]:

$$U_T = U_1 + U_2 = const. \tag{4}$$

That means:

$$dU_1 = -dU_2. \tag{5}$$

The entropy change in the subsystems can be evaluated according to equation (2) considering constraints (i) and (ii),

$$(dS)_1 = \frac{1}{T}\left(dU_1 - (\mu_A)_1(\delta N_A)_1 - (\mu_B)_1(\delta N_B)_1\right) = \frac{1}{T}\left[dU_1 - (\mu_A - \mu_B)_1(\delta N_A)_1\right], \tag{6a}$$

$$(dS)_2 = \frac{1}{T}\left(dU_2 - (\mu_A)_2(\delta N_A)_2 - (\mu_B)_2(\delta N_B)_2\right) = \frac{1}{T}\left[dU_2 - (\mu_A - \mu_B)_2(\delta N_A)_2\right]. \tag{6b}$$

The total entropy change is just the sum of (6a) and (6b), according to (iii) and (5) it results:

$$(dS)_T = \frac{1}{T}\left[(\mu_A - \mu_B)_R - (\mu_A - \mu_B)_G\right]\delta N_A. \tag{7}$$

In (7), making use of equation (iii), it is defined $\delta N_A = (\delta N_A)_1 = (\delta N_A)_2$. The application of equilibrium condition (3) to equation (7) results in the equilibrium condition for the chemical potentials of the exchanging ions in the binary ion exchange:

$$(\mu_A - \mu_B)_1 = (\mu_A - \mu_B)_2. \tag{8}$$

Equation (8) is the same condition for equilibrium proposed by Araujo[25] and it represents the equilibrium condition for binary Ion-Exchange.

It is worth noting that in most thermodynamic treatments of diathermal, rigid and permeable subsystems[25] or isothermal equilibria[24], the equilibrium condition is set by the result of the chemical potentials of each ion with the same value in the system and the reservoir. This last conclusion is not correct for binary Ion-Exchange involving two ions with constraints on extensive and corresponding conjugate variables.

### 3. Near surface thermodynamics of ion exchange

Chemical potential of ion $X$ in the reservoir or in the glass matrix is connected to its thermodynamic activity $a_X$ according to the following equation[26,29]:

$$\mu_X = \mu^0 + RT\ln(a_X). \tag{9}$$

Based on relationship (9), equation (8) can be written in terms of activities of ions $A$ and $B$ in the glass and in the reservoir as follows:

$$\frac{a_{\bar{A}}}{a_{\bar{B}}} \cdot \frac{a_B}{a_A} = K . \quad (10)$$

In equation (10) it has been introduced the equilibrium constant K which, according to (8), should be ideally equal to 1. Activities are related to concentrations of ions in the glass ($c_A, c_B$) and in the reservoir ($C_A, C_B$) by the introduction of the activity coefficients $\gamma$. For the reservoir, which is typically a liquid system consisting in molten salts, ion activities are expressed in terms of their concentrations in the reservoir and the respective activity coefficients ($\gamma_A, \gamma_B$):

$$\frac{a_B}{a_A} = \frac{\gamma_B C_B}{\gamma_A C_A} . \quad (11)$$

For the glass, which can be considered a solid system at the ion-exchange temperatures, it can be considered the classical approach based on activity coefficients. Another suitable approach for glass is the one suggested by Rothmund and Kornfeld[30]. Indicating with $c_A$ and $c_B$ the concentrations of ions A and B in the silicate glass it results:

$$\frac{a_{\bar{A}}}{a_{\bar{B}}} = \left(\frac{c_A}{c_B}\right)^n . \quad (12)$$

The above outlined approach to the discussion of Ion-Exchange equilibria conditions between glass and molten salts is the one proposed by Garfinkel[31]. Using positions (11) and (12) in equation (10) a final equation can be established in terms of ion concentrations in the reservoir (molten salt) ($C_A, C_B$) and in the glass ($c_A, c_B$):

$$\log\left(\frac{C_A}{C_B}\right) - \log\left(\frac{\gamma_B}{\gamma_A}\right) = n \log\left(\frac{c_A}{c_B}\right) - \log(K) \quad , \quad (13)$$

with "log" it is indicated the logarithm in base 10 while with "ln" the logarithm in base "$e$". The ratio of activity coefficients in molten nitrate mixtures can be approximated[31] considering regular simple symmetric mixtures[29,32] with Redlich-Kister expression limited to the first term[30]:

$$RT \ln(\gamma_A) = W_R(1-C_A)^2. \tag{14}$$

In the simple mixture theory, the constant "$W_R$" is the heat of mixing of the binary components of the mixture. Based on (14) and considering normalized relative concentrations: $C_A+C_B=1$ and $c_A+c_B=1$ and that the heat of mixing of alkali nitrates in the molten salt is not substantially depending on composition, the following expression, after logarithm's base change, can be obtained:

$$\log\left(\frac{\gamma_B}{\gamma_A}\right) = \frac{W_R}{2.303RT}(1-2C_B), \tag{15}$$

This position allows to write:

$$\log\left(\frac{C_A}{C_B}\right) - \frac{W_R}{2.303RT}(1-2C_B) = n\log\left(\frac{c_A}{c_B}\right) - \log(K). \tag{16}$$

To complete the set of equations for the discussion of thermodynamic equilibrium in Ion Exchange for silicate glass we can introduce the activity coefficients for ions $A$ and $B$ in the silicate glass matrix in a similar way as already defined for the reservoir:

$$\frac{a_{\bar{A}}}{a_{\bar{B}}} = \frac{\gamma_{\bar{A}} c_A}{\gamma_{\bar{B}} c_B}. \tag{17}$$

Following this argument in the same way as done for the reservoir an equation similar to (15) it can be set for the glass matrix:

$$\ln\left(\frac{\gamma_{\bar{A}}}{\gamma_{\bar{B}}}\right) = -\frac{W_{A/B}}{RT}(1-2c_B), \tag{18}$$

In this case $W_{A/B}$ can be interpretated as interaction energy of ions $A$ and $B$ in the glass matrix. The ratio of activity coefficients of ions $A$ and $B$ in the glass matrix can be expressed in terms of the coefficient $n$ introduced in equation (12):

$$\ln\left(\frac{\gamma_{\bar{A}}}{\gamma_{\bar{B}}}\right) = (n-1)\ln\left(\frac{c_A}{c_B}\right). \tag{19}$$

Taking the first term of the logarithm series expansion[31], ($\ln(z) \approx 2\left[\left(\frac{z-1}{z+1}\right)\right]$) and considering that ($c_A+c_B=1$) it results:

$$\ln\left(\frac{c_A}{c_B}\right) \approx 2(1-2c_B). \tag{20}$$

It is worth noting that the approximation of equation (20) is a good approximation for $c_B$ in the range 0.2-0.8 while it became critical as $c_B$ approaches 0 or 1.

Combining equation (18) and (19) with position (20) it is possible to express the coefficient "$n$" in terms of the interaction energy $W_{A/B}$ of neighboring $A$ and $B$ ions:

$$n = 1 - \frac{W_{A/B}}{2RT}. \tag{21}$$

From equation (21) it is evident that values of $n>1$ results in interaction energy $W_{A/B}<0$ that is a repulsive interaction energy of the exchanging ions in the glass matrix. When $n=1$ interaction energy is zero that means the behavior of exchanging ions in the glass matrix resembles a regular solution.

To complete the set of equations, at a zero-order approximation, activity coefficients can be set equal to 1 so that concentrations can take the place of activities. Following this further approximation and according to equation (10), the concentration ratio of the ions in the glass can be correlated to the same concentration ratio in the reservoir:

$$\frac{c_{\bar{A}}}{c_{\bar{B}}} \approx K \frac{c_A}{c_B}. \tag{22}$$

### 4. Review of past literature results

The first systematic experimental study of ion exchange equilibria between glass and molten salts is reported by Garfinkel[32] by using a direct determination method of the distribution of ions between the solution and the exchanger phase. In the Garfinkel study they have been used equilibrated powders. Molten salt mixtures were prepared ranging from 0.1 to 1 cation fraction of exchangeable species. Glass samples and molten salts have been equilibrated for 200-500 hours continuously stirring the mixtures. After equilibration glass samples and salts were separated and glass powders analyzed by flame photometry and titrimetry. Tests have been performed on several different alkali aluminosilicate glasses. They have been studied (Li/Na) exchange pairs at 400°C finding values of $n$=1.9 for an alkali-aluminosilicate glass (11.4 $Li_2O$·16.5$Al_2O_3$·71.5$SiO_2$ mol%) and $n$=3.2 for a glass with a lower content of alkali (5.86 $Li_2O$·11.5$Al_2O_3$·74.7$SiO_2$ mol%). Exchange pair (Na/K) has also been studied at 500°C for an alkali-alminosilicate glass (16.3 $Na_2O$·13.2$Al_2O_3$·66.7$SiO_2$ mol%) finding a value of $n$=1.2. In all studied cases it has been found $n$>1 values indicating repulsive energy of exchanging alkali in the glass matrix. Alkali Borosilicate glasses have been studied by Steyn and De Wet[34,35] in a wide range of temperatures (300°C-600°C) and for two different exchange situations: (Na/Li)$NO_3$ and (Na/K)$NO_3$ systems. In both cases it has been found[34] that $n$ values decrease as temperature increase where, in the (Na/Li) system, the value of $n$=1 is in between 400-450°C while for system (Na/K) is between 500-550°C. The change of $n$ from values above and below 1 indicates a change in the interaction energy between exchanging ions in the glass matrix from repulsive to attractive. In Figure 3 they are reported the "$n$" values as a function of temperature for the two exchange systems studied by Steyn and De Wet[34]. Dotted lines in Figure 3 are shown only for guidance of the trend of the values with temperature. The line $n$=1 is shown in the graph of figure 3 to identify the regular solutions behaviour.

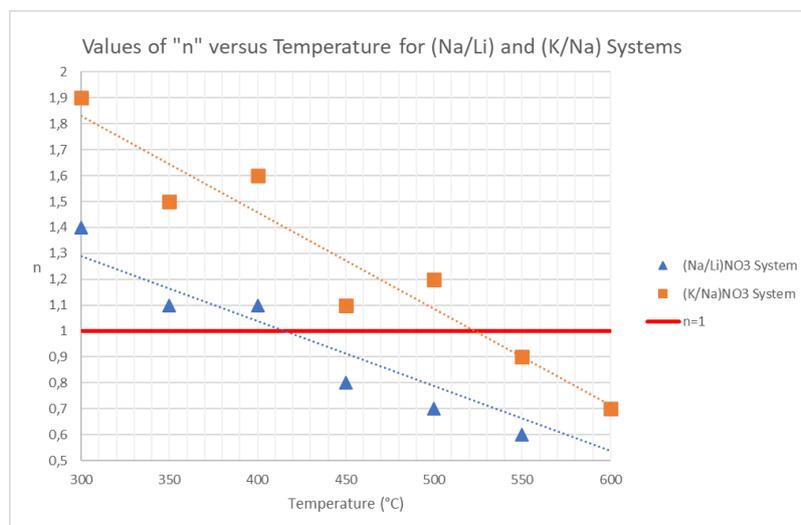

Figure 3 – Values of "$n$" (Equation (21)) versus exchange temperature for two different exchanging pairs in Alkali-Borosilicate glass (data from Steyn,DeWet[34]). Dotted lines are only for guidance.

A similar study has been reported by Orgaz and Navarro[36] using a determination technique similar to the one of Garfinkel. They have been studied again alkali aluminosilicate glasses with powders equilibrated at 350°C for 100-150 hours. They have been studied two glass series with a different content of Sodium oxide: a A-type glass (15 $Na_2O·15Al_2O_3·60SiO_2$ mol%) and a B-type glass (10 $Na_2O·15Al_2O_3·60SiO_2$ mol%). Values of n have been found with values ranging from $n=2.3$ to $n=1.9$ for A-Type and $n=3.6$ to $n=2.5$ for B-Type glass. It is worth mentioning a more recent experimental study reported by Patschger and Russel[37], They have been studied Soda-lime glass (13.6 $Na_2O·1.0Al_2O_3·71.1SiO_2$ mol%) and aluminosilicate glass (14.1 $Na_2O·9.0Al_2O_3·68.2SiO_2$ mol%). It is interesting to point out that the determinations have been performed in a direct ion exchange process performed at 470°C for 6 hours on samples slabs cut from flat parts. They have been determined Sodium and Potassium surface concentration of the ion exchanged samples by SEM with EDAX system and EDX for concentration depth profiles. Patschger and Russel report a value of $n=1.19$ for soda-lime glass and $n=1.75$ for sodium-aluminosilicate glass. Again, a "*n*" value above 1 indicates repulsive interaction energy of exchanging ion pair in the glass matrix. The value of the equilibrium constant in the Patschger and Russel study[37] for ion exchange at 470°C for soda-lime glass $K$(S/L)=0.13 and sodium-aluminosilicate glass $K$(ALS)=0.47. This results allow to determine the concentration ratio of exchanging ions on the glass surface as a function of the same concentration ratio in the molten salt through equation (22). In figure 4 the [Na]/[K] ratio at glass surface is reported as a function of the same ratio in the molten salt calculated from equation (22) using the equilibrium constant values determined by Patschger and Russel[37]. During ion exchange the molten salt chemical composition is altered by the enrichment of Sodium ions coming from the glass. This may have a significant effect to the surface concentration of Potassium ions. A result is presented by Hale[38] where it has been evaluated the effect of mole fraction of Sodium ion in molten salt to the efficiency of exchange at the glass surface.

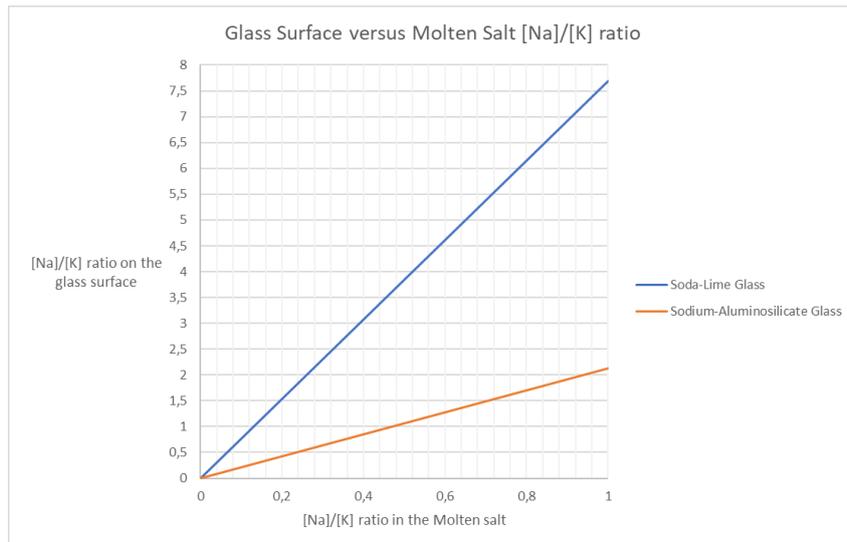

Figure 4 – [Na]/[K] ion exchange ratio at glass surface as a function of [Na]/[K] in the molten salt. (Data for equilibrium constant from Patscher and Russel[37].

In the Hale study the equilibrium constant $K$ is evaluated considering the mechanical work involved in the replacement of one mole of Sodium ions in the glass by an equimolar amount of Potassium ions coming from the molten salt. This evaluation is performed by using the Eshelby misfitting sphere theory. In Figure 5 it is presented the surface exchange percentage values coming from the Hale theory[38] with the ones of Patschger and Russel[37].

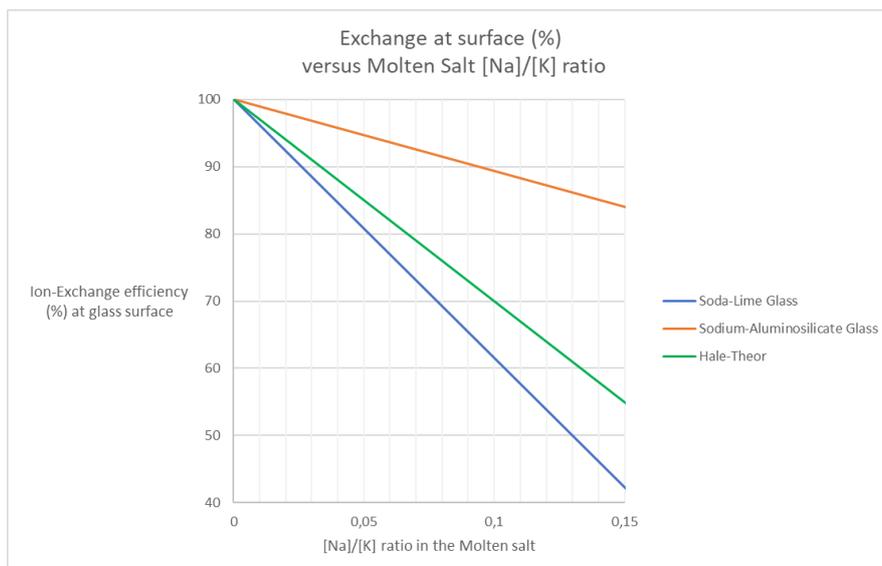

Figure 5 – Ion Exchange efficiency (%) as a function of [Na]/[K] ratio in the molten salt (Data for Soda-Lime and Sodium-Aluminosilicate glass from Patschger and Russel[37] and theoretical data from Hale[38].

The Hale curve has been calculated considering a theoretical equilibrium constant value of $K_{Hale}$=0.17. From Figures 4 and 5 it is evident how for Soda-lime glass a limited increase in the Sodium concentration in the molten salt significantly reduce the exchange efficiency at glass surface. From figure 5 it results that with a ratio [Na]/[K] in the molten salt of 0.1 the surface exchange efficiency is reduced at 60%. This result provides a technological indication to mitigate the effects of Sodium increase in the molten salt that can reduce the surface exchange efficiency. This can be achieved by having an excess of Potassium Nitrate in comparison to glass volume. The same result is reported also by Kirchner and Mauro[39] where they consider the need to increase salt bath-to-glass volume ration ($Vb/Vg$) in order to maximize entropy production which is the driving thermodynamic force for ion exchange. It is significant to connect this last argument to the thermodynamic treatment of ion-exchange equilibria where the equilibrium condition (Equation (8)) has been derived under the condition to maximize the total entropy of the "molten salt/glass system".

## 5. Kinetics of ion exchange

Kinetics of ion exchange is widely discussed in the literature[40,41]. It is clearly a non-equilibrium interdiffusion process where the fluxes of the exchanged ions are related to the gradient of the respective electrochemical potentials and they are subject to suitable constraints dictated by Gibbs-Duhem equation[26,27,28] and the conservation of overall fluxes. In a first order approximation the influence of residual stress can be neglected[41] and, in mathematical terms, the flux equations are:

$$-J_i = \beta_i c_i \left( \frac{\partial \mu_i}{\partial x} - FE \right); i=A,B, \qquad (23)$$

where $\beta_i$ is the mobility of the ions in the matrix, $E$ is the electric field and $F$ the Faraday constant. We can express mobility through the self-diffusion coefficient $D_i$ according to the Einstein equation:

$$D_i = \beta_i RT . \qquad (24)$$

The kinetic equations and constraints are:

$$-J_A = \frac{D_A}{RT} c_A \left( \frac{\partial \mu_A}{\partial x} - FE \right), \quad (25\text{i})$$

$$-J_B = \frac{D_B}{RT} c_B \left( \frac{\partial \mu_B}{\partial x} - FE \right), \quad (25\text{ii})$$

$$J_A + J_B = 0, \quad (25\text{iii})$$

$$c_A d\mu_A + c_B d\mu_B = 0. \quad (25\text{iv})$$

Application of conditions (25iii) and (25iv) to flux equations (25i) and (25ii) allows to calculate the electric field generated by the difference of the ion mobilities in the glass matrix:

$$FE = \frac{D_A - D_B}{c_A D_A + c_B D_B} c_A \frac{\partial \mu_A}{\partial x}. \quad (26)$$

The insertion of (26) into thet flux equation (25i) results:

$$-J_A = \frac{c_A}{RT} \left[ \frac{D_A D_B (c_A + c_B)}{c_A D_A + c_B D_B} \right] \frac{\partial \mu_A}{\partial x}, \quad (27)$$

making use of equation (9) which connect chemical potential to activity and introducing the relative concentrations:

$$\chi_A = \frac{c_A}{c_A + c_B}; \chi_B = \frac{c_B}{c_A + c_B}, \quad (28)$$

We came to the flux equation:

$$-J_A = \left[ \frac{D_A D_B}{\chi_A D_A + \chi_B D_B} \frac{\partial \ln a_A}{\partial \ln c_A} \right] \frac{\partial c_A}{\partial x}. \quad (29)$$

Defining an interdiffusion coefficient:

$$\tilde{D}_{AB} = \frac{D_A D_B}{\chi_A D_A + \chi_B D_B} \frac{\partial \ln a_A}{\partial \ln c_A}, \quad (30)$$

We can write a familiar expression for the flux equation:

$$-J_A = \tilde{D}_{AB} \frac{\partial c_A}{\partial x}. \quad (31)$$

The activity derivative in equation (30) is known as "thermodynamic factor" and it can be expressed in two different forms depending on the expression of activity:

$$a_A = \gamma_A c_A, \tag{32i}$$

$$\tilde{D}_{AB} = \frac{D_A D_B}{\chi_A D_A + \chi_B D_B} \frac{\partial \ln a_A}{\partial \ln c_A} = \frac{D_A D_B}{\chi_A D_A + \chi_B D_B}\left(1 + \frac{\partial \ln \gamma_A}{\partial \ln c_A}\right). \tag{32ii}$$

Or, following Rothmund and Kornfeld[30]:

$$a_A = (c_A)^n, \tag{33i}$$

$$\tilde{D}_{AB} = \frac{D_A D_B}{\chi_A D_A + \chi_B D_B} \frac{\partial \ln a_A}{\partial \ln c_A} = \frac{D_A D_B}{\chi_A D_A + \chi_B D_B} n. \tag{33ii}$$

In the literature[33], it is defined the thermodynamic factor $n$:

$$n = \frac{\partial \ln a_A}{\partial \ln c_A} = 1 + \frac{\partial \ln \gamma_A}{\partial \ln c_A}. \tag{34}$$

When $n=1$ the kinetic behavior is called a regular solution behavior. It has been demonstrated[41] that the thermodynamic factor "$n$" is of critical relevance to reconcile experimental results for the interdiffusion coefficient. Using experimental data of Varshneya[42] for ion exchange of soda-lime glass at 350°C, neglecting both the thermodynamic factor ($n=1$) and stress effects, the interdiffusion coefficient $\tilde{D}_{AB}$ (33ii), presents critical matching at the surface ($\chi_K=1$) and towards the bulk ($\chi_K=0$) as shown in Figure 6.

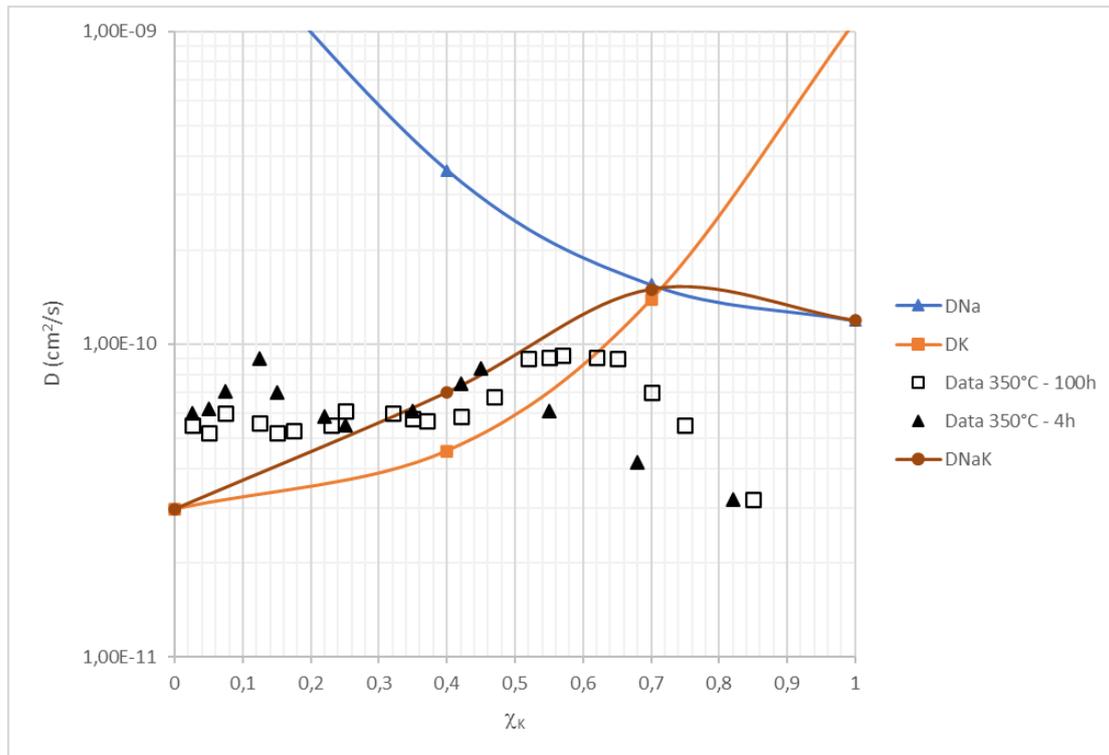

Figure 6 – Interdiffusion coefficient comparison with experimental data from Varshneya[42] at 350°C, $n$=1 (no effect of thermodynamic factor $n$=1, no effect of stress).

To evaluate the effect of the thermodynamic factor and residual stress introduced by ion exchange[41], the n value at 350°C has been considered based on the results of Patschger and Russel at 470°C ($n$=1.2) and the trend of figure 3. A reasonable value is set at $n$=1.5. The resulting curve for this *Deff* ($n$=1.5, Stress effects) is reported in Figure 7.

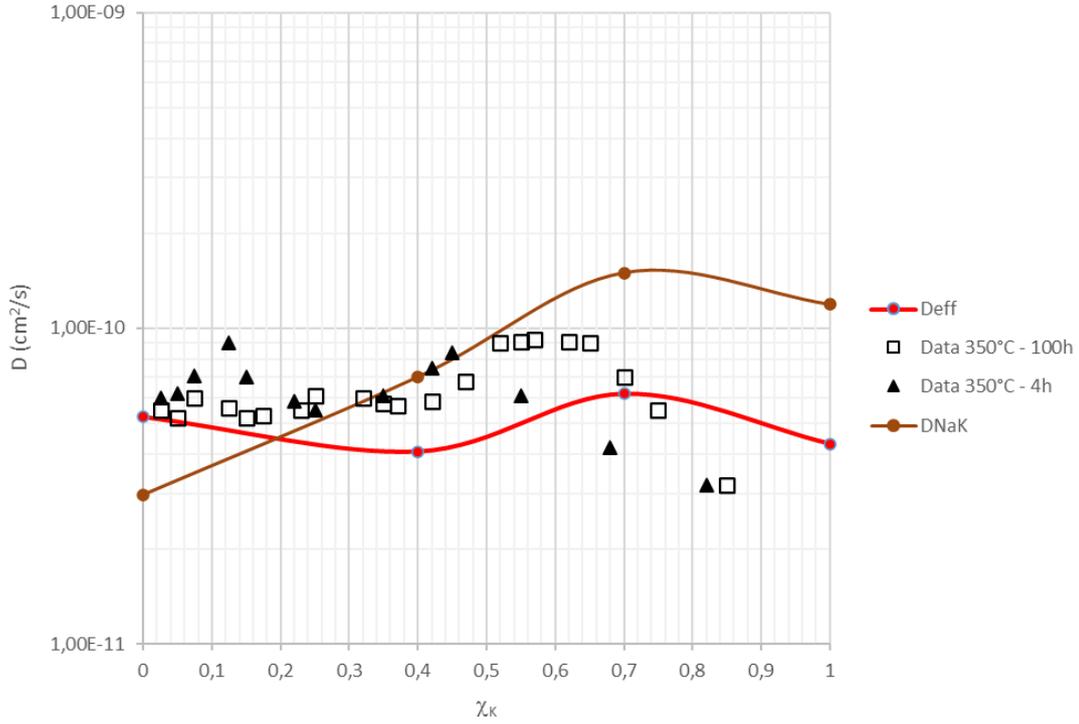

Figure 7 – Interdiffusion coefficient comparison, experimental data from Varshneya[42] at 350°C, $D_{NaK}$ (brown curve) evaluated with $n=1$ (no effect of thermodynamic factor n=1, no effect of stress), $Deff$ (red curve) evaluated with $n=1.5$ and hydrostatic residual stress effects[41].

It is clear from Figure 7 that surface and bulk data are better represented considering both the residual hydrostatic stress and thermodynamic factor influence.

The purpose of the kinetics of ion exchange[14,18] is the determination of the concentration of the incoming ions in the glass $c_A(x,t)$. This means being able to write a diffusion equation for the concentration $c_A(x,t)$ that, under suitable boundary conditions, can be solved by analytical or numerical techniques. The starting point to write down a diffusion equation is to consider a mass conservation equation for the ion flux[43]:

$$\frac{\partial c_A(x,t)}{\partial t} + \frac{\partial J(x,t)}{\partial x} = 0, \qquad (35)$$

coupled with a so-called constitutive equation as a Fick type equation (31) leading to:

$$\frac{\partial c_A(x,t)}{\partial t} - \frac{\partial}{\partial x}\left(\tilde{D}_{AB}\frac{\partial c}{\partial x}\right) = 0. \qquad (36)$$

Assuming, as first order approximation, a constant interdiffusion coefficient[14,18] the diffusion equation results in the familiar form:

$$\frac{\partial c_A(x,t)}{\partial t} - \tilde{D}_{AB}\frac{\partial^2 c}{\partial x^2} = 0. \tag{37}$$

The diffusion equation (36) is a second-order partial derivative differential equation, its solution (apart from the mathematical conditions for the existence of a solution) requires boundary conditions. The typical boundary condition considered is related to the initial assumption we have already taken that the equilibrium condition at the interface Molten Salt/Glass is achieved in a much faster time than the overall kinetic process. This means that we can consider at time zero of the process a constant value of surface concentration that remain constant for the entire duration of the kinetic process:

$$c_A(0,t) = c_{sA}; \text{ when } t \geq 0. \tag{38}$$

Condition (38) is known in mathematics as first kind boundary condition. The solution is well known in the literature[14,18,43]:

$$c_A(x,t) = c_{sA} \cdot erfc\left(\frac{x}{2\sqrt{\tilde{D}_{AB} \cdot t}}\right), \tag{39}$$

where the *erfc(z)* is the well-known[31,43] complementary error function:

$$erfc(z) = 1 - erf(z) = 1 - \frac{2}{\sqrt{\pi}} \cdot \int_0^z e^{-\eta^2} d\eta. \tag{40}$$

The concentration in the glass is finally used for the determination of the residual stress generated by ion-exchange through the expressions given by Cooper which neglect relaxation effects[14,18]:

$$\sigma(x,t) = -\frac{B \cdot E}{1-\nu}\left[c_A(x,t) - \overline{c}_A(t)\right], \tag{41}$$

where *E* is the Young modulus of the glass, $\nu$ its Poisson's ratio, and *B* is the linear network dilation coefficient, also known as the Cooper coefficient. In case stress relaxation effects are relevant equation (41) becomes[17,19]:

$$\sigma(x,t) = -\frac{B \cdot E}{1-\nu}\left[\left[V(x,t)c_A(x,t) - \langle Vc_A(t)\rangle\right] - \int_0^t \frac{\partial R(t-\theta)}{\partial \theta}\left[V(x,\theta)c_A(x,\theta) - \langle Vc_A(\theta)\rangle\right]d\theta\right], \tag{42}$$

where $R(t)$ is the viscous relaxation function and $V(x,t)$ is the Varshneya function that takes into account fast $\beta$ and slow $\alpha$ structural relaxation mechanisms[17,19]. From what discussed it is clear the role of surface concentration in the determination of the incoming ions concentration. Surface concentrations of the incoming ($A$) and outgoing ($B$) ions in the glass are related to the concentration of the same ions in the molten salt through the equilibrium isotherm parameters $K$ and $n$ through the following equation which is derived from equation (10):

$$\frac{c_{s\bar{A}}}{c_{s\bar{B}}} = \left[ K \frac{\gamma_A c_A}{\gamma_B c_B} \right]^{1/n}, \qquad (43)$$

where $c_{sA}$ and $c_{sB}$ are the ions concentration at the glass surface. Under the assumption of $\gamma_A = \gamma_B$ and $n=1$ equation (43) reduces to equation (22). It is relevant here to recall the experimental study of Patschger and Russel[37] where they have measured surface concentration of Sodium and Potassium for two types of glasses: Soda Lime Silicate (SLS) and Sodium-Aluminosilicate (SAS) submitted to ion exchange (6hours at 470°C) with molten salt baths of mixed $KNO_3$ and $NaNO_3$ with different ratio of composition (see Table 1).

Table 1 – Composition of molten salt baths of the Patschger and Russel study[37]

| Batch ID | KNO3 (%) | NaNO3 (%) | [Na]/[K] |
|---|---|---|---|
| 1 | 100 | 0 | 0,00 |
| 2 | 90 | 10 | 0,11 |
| 3 | 80 | 20 | 0,25 |
| 4 | 70 | 30 | 0,43 |
| 5 | 60 | 40 | 0,67 |
| 6 | 50 | 50 | 1,00 |

In Figure 8 they are reported the ratio of surface concentration of Sodium and Potassium as a function of the same ratio in the different salt baths. The data points and the linear interpolation dotted curves indicate that the assumptions leading to equation (22) and linear curves as the ones reported in Figure 4 are consistent. The equilibrium constant $K$ is dominating in determining the equilibrium surface concentration. The effect on kinetics, namely the Potassium concentration depth profiles, has been determined experimentally by Patschger and Russel[37] by performing a depth profile analysis.

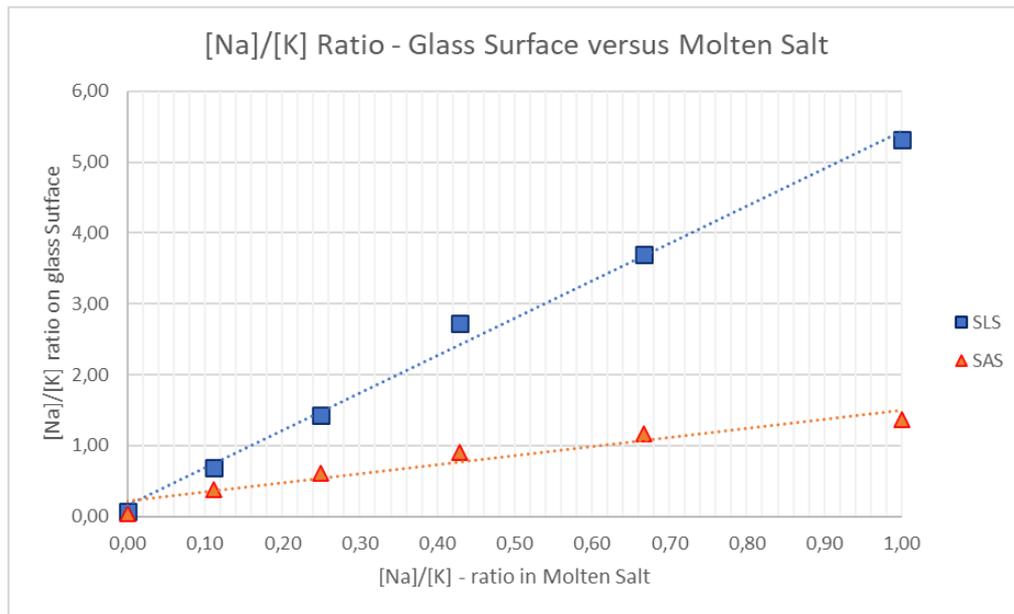

Figure 8 – [Na]/[K] ratio on glass surface versus [Na]/[K] ratio in the molten salts.

The assumption of constant interdiffusion coefficient has been found consistent, values are reported in Table 2 for both types of glasses and concentration profiles according to equation (39) are presented in Figure 9.

Table 2 – Constant Interdiffusion coefficients for SAS and SAS Glass for the different molten salt baths data from Patschger and Russel study[37]

| Molten salt bath ID | Interdiffusion Coefficient $D_{Na/K}$ [cm2/s] | |
|---|---|---|
| | SLS | SAS |
| 1 | 2,20E-11 | 5,10E-10 |
| 2 | 2,60E-11 | 6,60E-10 |
| 3 | 1,60E-11 | 5,00E-10 |
| 4 | 1,90E-11 | 5,70E-10 |
| 5 | 2,20E-11 | 5,70E-10 |
| 6 | 1,70E-11 | 5,10E-10 |

Concentration profiles are calculated according to equation (39) taking the correspondent surface concentration equilibrium values and they are shown in Figure 9 for both SLS and SAS glasses.

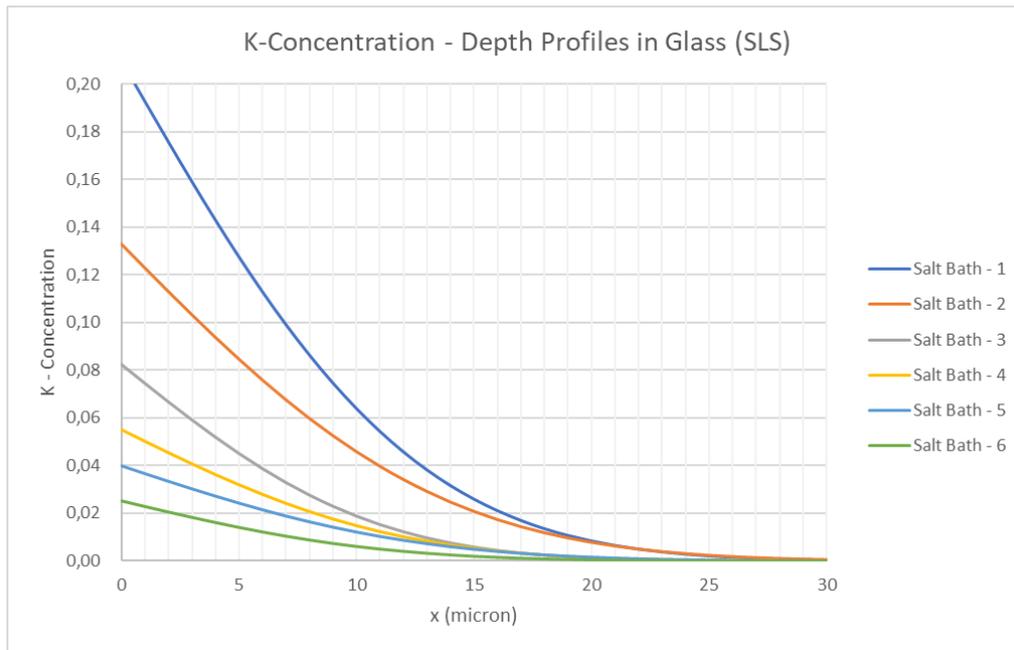

a)

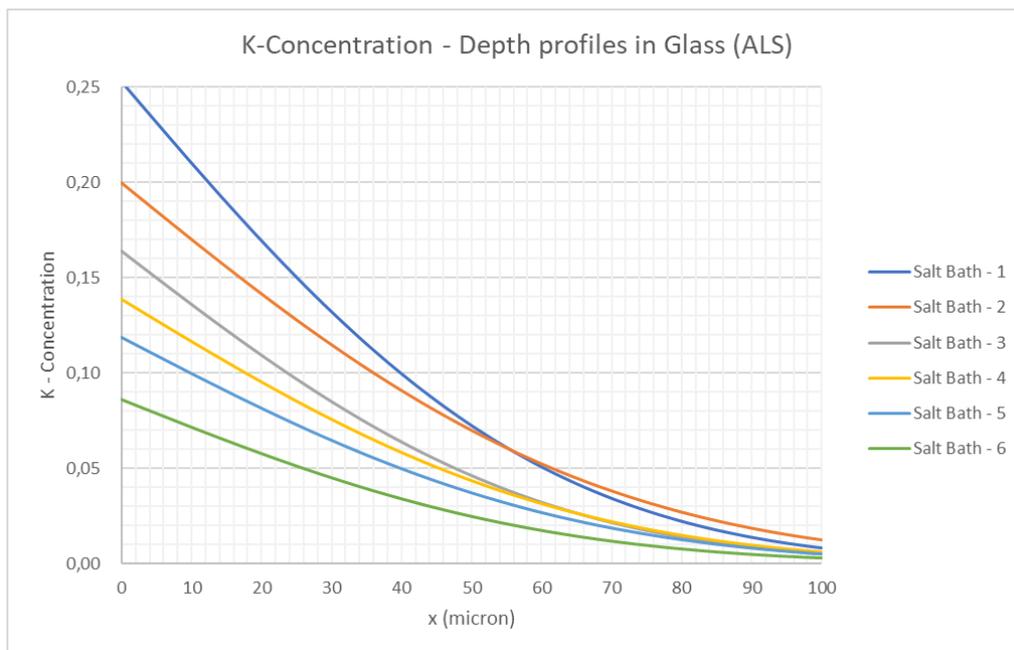

b)

Figure 9 – Depth profile concentrations for glass SLS a) and SAS b), calculated according to equation (39) using the interdiffusion coefficient values of Table 2 and surface concentration of Figure 8 – Experimental data from Patschger and Russel[37]. Data of concentration are expressed as normalized values (see cited reference for details).

The concentration values at the surface ($x=0$) in Figure 9 represent the surface concentration equilibrium values for the different baths.

## 6. Ion exchange in the framework of non-equilibrium thermodynamics.

The kinetics flux equations (23) represent an approximation of a more general expression of the ions fluxes[18]. As already pointed out in this study, kinetics phenomena like ion-exchange are non-equilibrium processes that can be approached within a non-equilibrium thermodynamic theory[27,44]. The extension of the discussion about ion-exchange will be herewith limited to the linear thermodynamics of irreversible processes[18,44]. They can be introduced the affinities $F_i$ as the thermodynamic driving forces of an irreversible process in terms of entropy increase of the universe (overall surrounding of the system) due to the changing of the corresponding kinetics thermodynamic coordinates $X_i$:

$$F_i = \frac{\partial S}{\partial X_i}. \tag{44}$$

The second definition is related to the response of the system to the affinities that are the fluxes $J_i$ which are defined by the time rate of change of thermodynamic coordinates $X_i$:

$$J_i = \frac{\partial X_i}{\partial t}. \tag{45}$$

The definitions of affinities and fluxes allow the expression of the increase of entropy of the universe as result of irreversible processes[18,44]

$$\frac{dS}{dt} = \sum_i \frac{\partial S}{\partial X_i} \frac{\partial X_i}{\partial t} = \sum_i F_i J_i. \tag{46}$$

The linear thermodynamic of irreversible processes is based on the assumption that the fluxes, at a defined point in time, depend only on the affinities at the same point in time and not on any previous state of the system[18]. This assumption allows the expansion of fluxes in powers of the affinities by a Taylor series approximation that become a MacLaurin series because fluxes are zero when affinity are null[18]. An additional assumption is to consider the affinities reasonably small such that quadratic and higher terms in the MacLaurin series can be neglected and a final expression of fluxes can be written in a linear form:

$$J_i = \sum_i L_{ij} F_j. \qquad (47)$$

In a binary ion exchange process the equations system (47) can be explicitly written:

$$J_A = L_{AA} F_A + L_{AB} F_B \quad, \qquad (48\text{i})$$

$$J_B = L_{BA} F_A + L_{BB} F_B \quad. \qquad (48\text{ii})$$

The driving forces can be expressed in terms of chemical potentials ($\mu_A$ and $\mu_B$) of the exchanging ions and of the electric field $E$ due to their different mobilities.

$$F_A = -\frac{\partial \mu_A}{\partial x} + q\mathcal{F} E \quad, \qquad (49\text{i})$$

$$F_B = -\frac{\partial \mu_B}{\partial x} + q\mathcal{F} E \quad. \qquad (49\text{ii})$$

In equations (49) $q$ is the electric charge of exchanged ions (in our case $q=1$ because we are considering monovalent ions) and $\mathcal{F}$ is the Faraday constant. The equations (48i) and (48ii) with (49i) and (49ii) subjected to the flux conservation condition (25iii) and Gibbs-Duhem equation (25iv) have been discussed by Poling and Houde-Walter[45,46]. As correctly pointed out by Poling and Houde-Walter[45] following an argument already discussed by De Groot[47], the Onsager reciprocal relations $L_{ij}=L_{ji}$ are meaningless in this discussion because this mutual-diffusion depends from a single kinetic parameter. A similar discussion has been carried out by Tagantsev and Ivanenko[48] where they used in their appendix A the Onsager relationships coming to a similar general diffusion equation reached by De Groot[47] without using Onsager relationships. We can repeat the derivation of a flux equation like the (29) following the same approach outlined in paragraph 5. by using the above defined equations and conditions (conservation of fluxes and Gibbs-Duhem equation). After a lengthy but straightforward calculation considering (28) the final flux equation results:

$$-J_A = \left[ \frac{RT}{c_T} \frac{1}{\chi_A \chi_B} \left( \frac{L_{AA} L_{BB} - L_{AB} L_{BA}}{L_{AA} + L_{BB} + L_{AB} + L_{BA}} \right) \right] n \frac{\partial c_A}{\partial x} \quad, \qquad (50)$$

where $n$ is the already introduced thermodynamic factor (34), and $c_T = c_A + c_B$. The diffusion coefficients can be conveniently introduced[45]:

$$D_{ij} = \frac{RT}{c_T} \frac{L_{ij}}{\chi_j} \quad \text{for i = A,B (i} \neq \text{j);} \quad D_i = \frac{RT}{c_T} \frac{L_{ii}}{\chi_i} \quad \text{for i = A,B} \ . \tag{51}$$

With this definitions equation the flux equation (50) can be written in terms of diffusion coefficients:

$$-J_A = \frac{D_A D_B - D_{AB} D_{BA}}{\chi_A (D_A + D_{BA}) + \chi_B (D_B + D_{AB})} n \frac{\partial c_A}{\partial x} \ . \tag{52}$$

Incidentally the above flux equation is quite similar to the one derived by Poling and Houde-Walter[45]. The next step is to try a factorization of the interdiffusion coefficient of equation (52) in order to make a direct comparison with flux equation (29) derived without taking into account the cross-diffusion terms resulting from the linear irreversible thermodynamics. To develop such a factorization let's define the interdiffusion coefficient of equation (52) as follows:

$$D_{AB}^* = \frac{D_A D_B - D_{AB} D_{BA}}{\chi_A D_A + \chi_B D_B + \chi_A D_{BA} + \chi_B D_{AB}} n \ . \tag{53}$$

The interdiffusion coefficient (53) can be factorized in terms of the interdiffusion coefficient (30) and (33ii)

$$D_{AB}^* = \tilde{D}_{AB} \cdot \Psi_{Cross}, \tag{54}$$

The explicit calculation of $\Psi_{cross}$ is straightforward and it results:

$$\Psi_{cross}(\chi_A, \chi_B) = \frac{\chi_A D_A + \chi_B D_B}{\chi_A D_A + \chi_B D_B + \chi_A D_{BA} + \chi_B D_{AB}} \left(1 - \frac{D_{AB} D_{BA}}{D_A D_B}\right). \tag{55}$$

Finally, equation (52) can be written as a generalization of equation (31):

$$-J_A = \tilde{D}_{AB} \Psi_{cross} \frac{\partial c_A}{\partial x} \ . \tag{56}$$

Equation (56) allows the introduction of a new thermodynamic factor $\psi(\gamma_A, \gamma_B)$ which is a function of glass chemical composition according to:

$$\psi(\gamma_A, \gamma_B) = \Psi_{cross}(\gamma_A, \gamma_B) \cdot n \tag{57}$$

As correctly pointed out by Poling and Houde-Walter[45] the cross-terms in flux expressions introduced with the generalization to linear irreversible thermodynamics represent the interaction between the unlike species and, as such, they shall be introduced in the thermodynamic factor as we have suggested with equation (57).

## 7. Discussion

The relevance of the thermodynamic factor "$n$" and of the equilibrium constant "$K$" to the description of ion exchange kinetics is clearly evident. Thermodynamic equilibrium analysis allows the determination of "$n$" to match in an acceptable way results coming from the determination of the interdiffusion coefficient while the equilibrium constant "$K$" is critical in the definition of both surface concentration and depth profile concentration. The effect of the enrichment of the ion reservoir (molten salt bath) during ion exchange by ions B coming from the glass can be conveniently modeled using the equilibrium constant "$K$" and it is reflected in the values of equilibrium surface concentration. This last effect is generally considered a pollution of the molten salt bath and it is of significant practical relevance in the technological application because of its influence on equilibrium surface concentration and, in turn, surface compression (when the process is used for glass strengthening and surface refractive index when the process is used to create optical waveguides). A potential further improvement of the theory is to consider $n$ no more constant but depending from the ion concentration. The concentration depending contribution of the thermodynamic factor to the determination of the interdiffusion coefficient can be considered on the basis of the result achieved with the extension of the ion-exchange kinetic theory of paragraph 5 to the linear irreversible thermodynamics approach outlined in paragraph 6. The interdiffusion coefficient corrected according to the linear irreversible thermodynamics theory is:

$$D^*_{NaK} = \frac{D_K D_{Na}}{\chi_K D_K + \chi_{Na} D_{Na}} \psi(\chi_K).$$

(58)

The corrected interdiffusion coefficient according to (58) is calculated in Figure 10 and compared with experimental data[42] considering $n=1.5$ at the glass surface and $n=2$ towards the bulk.

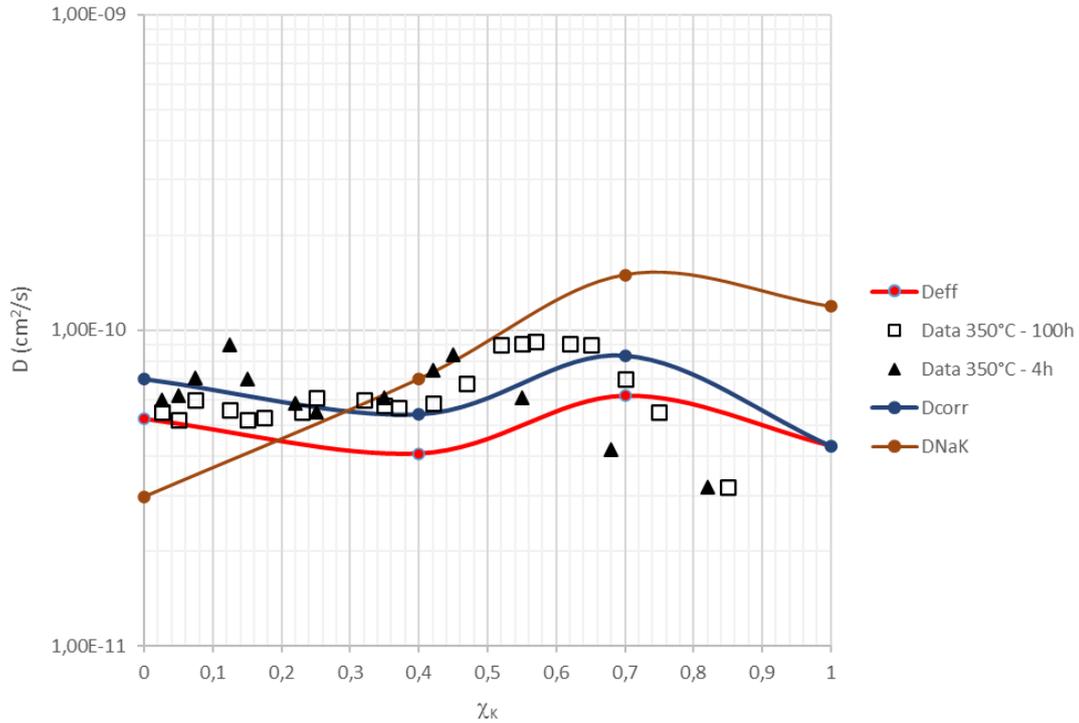

Figure 10 – Interdiffusion coefficient comparison, experimental data from Varshneya[42] at 350°C, $D_{NaK}$ (brown curve) evaluated with $n=1$ (no effect of thermodynamic factor $n=1$, no effect of stress), $D_{eff}$ (red curve) evaluated with $n=1.5$ and hydrostatic residual stress effects[41] $D_{corr}=D^*_{NaK}$ (blue curve) evaluated according to equation (58) setting $\psi=1.5$ at the surface and $\psi=2$ towards the bulk.

Looking to equation (22) $n$ is strictly connected to the interaction energy of the exchanging ions pair in the glass matrix. With the chosen values at 350°C we have that: $n=1.5$ corresponds to $W_{NaK} = -6.4$ kJ/mol while for n=2.0 we have $W_{NaK}= -12.8$ kJ/mol. This approach, in some way, overpasses the hypothesis of a further suggested[41] interaction of the exchanging ions with the silicon dioxide network. It can be argued that the two approaches can be reconciled by a further investigation demonstrating the composition dependence of $n$ due to an interaction of the ions with the network. This last issue has not been investigated in this study apart from the generalization of the thermodynamic factor by the introduction of cross terms effects to justify its concentration dependance. On the other side it shall be recognized that the use of constant interdiffusion coefficient is an acceptable approximation for technological applications where concentration is relevant to predict residual stress or refractive index profiles.

## 8. Conclusion

Equilibrium conditions for binary ion exchange in silicate glasses has been derived within the framework of the thermodynamic of subsystems. The condition is expressed in terms of the chemical potentials of the exchanged components and it results that equilibrium is achieved when the difference between chemical potentials of the two components is the same in the subsystems. This condition allows the introduction of an equilibrium constant "$K$" for the ion exchange chemical reaction. Considering the relationships between chemical potentials and concentrations through activities coefficients, it is possible to determine the isotherms of the ion exchange process. The development of the thermodynamic theory allows the introduction of the parameter "$n$" connected with the interaction energy of the exchanging ions pair in the glass matrix. The value of $n=1$ indicates a regular solution behavior versus a non-ideal behavior when $n \neq 1$. A review of past literature has substantially confirmed the possibility to determine the equilibrium isotherm in a direct way either through long term equilibration of glass powders in glass melts[32] or by the determination of surface and depth profile concentration[37] of glass samples submitted to ion exchange. The development of a kinetic theory of ion exchange clarifies the influence of both the thermodynamic factor "n" to the interdiffusion coefficient and of the equilibrium constant "$K$" to surface concentration and, in turn, to the final concentration distribution in the glass. A good agreement between past experimental results and calculated results for the interdiffusion coefficient obtained through the presented theory is obtained considering both the effects of introduced residual stress and the thermodynamic factor. Even better matching between experimental and theoretical results is achieved considering a concentration dependent thermodynamic factor $\psi(\chi_K)$ which has been justified by extending the ion-exchange kinetic theory to linear irreversible thermodynamics. The influence of the equilibrium constant "$K$" on the surface equilibrium concentration allows the determination of concentration profiles of the incoming ions in the glass assuming a simplified kinetic model with a constant interdiffusion coefficient. In this study it has been demonstrated a deep connection between equilibrium thermodynamic and kinetic analysis of ion exchange processes in silicate glass.